\documentclass[a4paper,10pt,english]{article}
\usepackage{graphicx, amsmath, bbold, tensor,xcolor} 
\usepackage{amsmath}

\usepackage{verbatim}
\usepackage{amssymb}
\usepackage{dynkin-diagrams, braket}
\usepackage{tikz}
\usetikzlibrary{calc, arrows.meta,shapes,positioning}
\usepackage[maxbibnames=99,sorting=none,style=numeric,citestyle=numeric-comp,backend=biber]{biblatex}
\usepackage{caption}
\usepackage[T1]{fontenc}

\DeclareFieldFormat{pages}{#1}
\renewbibmacro{in:}{}
\usepackage[colorlinks, allcolors=black]{hyperref}
\bibliography{bibliography} 

\begin{document}

\title{Unification of Gravity and Internal Interactions}
\author{Spyros Konitopoulos$^1$, Danai Roumelioti$^2$, George Zoupanos$^{2,3,4,5}$}
\date{2023}
\maketitle

\begin{center}
\itshape$^1$Institute of Nuclear and Particle Physics, NCSR Demokritos, Athens, Greece \\
\itshape$^2$Physics Department, National Technical University, Athens, Greece\\
\itshape$^3$Theory Department, CERN\\
\itshape$^4$Max-Planck Institut f\"ur Physik, M\"unchen, Germany\\
\itshape$^5$Institut f\"ur Theoretische Physik der Universit\"at Heidelberg, Germany
\end{center}

\begin{center}
\emph{E-mails: \href{mailto:spykoni@gmail.com}{{\color{black}spykoni@gmail.com}}, \href{mailto:danai\_roumelioti@mail.ntua.gr}{{\color{black}danai\_roumelioti@mail.ntua.gr}}, \\ \href{mailto:George.Zoupanos@cern.ch}{{\color{black}George.Zoupanos@cern.ch}}}
\end{center}

\begin{abstract}
In the gauge theoretic approach of gravity, General Relativity is
described by gauging the symmetry of the tangent manifold in four
dimensions. Usually the dimension of the tangent space is considered to
be equal to the dimension of the curved manifold. However, the tangent group of a manifold of dimension $d$ is not necessarily $SO_d$. It has been suggested earlier that by gauging an enlarged symmetry
of the tangent space in four dimensions one could unify gravity with
internal interactions. Here we consider such a unified model by gauging the $SO_{(1,17)}$ as the extended
Lorentz group overcoming in this way some difficulties of the previous attempts of
similar unification and eventually we obtain the $SO_{10}$ GUT, supplemented
by an $SU_2 \times SU_2$ global symmetry.
\end{abstract}
\newpage
\section{Introduction}
An ultimate aim of many theoretical physicists is the existence of a
unification picture in which all known fundamental interactions are involved.
A huge amount of serious research activity has been carried
out, including works that elaborate the very interesting notion of extra
dimensions. The earliest unification attempts of Kaluza and Klein
\cite{Kaluza:1921,Klein:1926} included gravity and electromagnetism, which were the established
interactions at that time. The proposal was to reduce a pure gravity theory from
five dimensions to four, which led to a $U_1$ gauge theory, identified with
electromagnetism, coupled to gravity. A revival of interest in the Kaluza-Klein
scheme started after realizing \cite{Kerner:1968, CHO1987358, PhysRevD.12.1711} that non-abelian gauge groups appear
naturally when one further extends the spacetime dimensions. With the assumption
that the total spacetime manifold can be written as a direct product $M_D =M_4\times B$,
where $B$ is a compact Riemannian space with a non-abelian isometry group $S$,
dimensional reduction of the theory leads to gravity coupled to a Yang-Mills
theory with a gauge group containing $S$ and scalars in four dimensions. The main
advantage of this picture is the geometrical unification of gravity with the
other interactions and also the explanation of gauge symmetries. There exist
serious problems though in the Kaluza-Klein framework, e.g. there is no
classical ground state corresponding to the direct product structure of $M_D$. However, the
most serious obstacle in obtaining a realistic model of the low-energy
interactions seems to be that after adding fermions to the original action it is
impossible to obtain chiral fermions in four dimensions \cite{Witten:1983}. Eventually, if one adds
suitable matter fields to the original gravity action in particular Yang-Mills
then most of the serious problems are resolved. Therefore one is led to
introduce Yang-Mills fields in higher dimensions. In case the Yang-Mills are
part of a Grand Unified Theory (GUT) together with a Dirac one \cite{PhysRevLett.32.438, FRITZSCH1975193}, the
restriction to obtain chiral fermions in four dimensions is limited to the
 requirement that the total dimension of spacetime should be $4k+2$ (see e.g.
\cite{CHAPLINE1982461}). During the last decades the Superstring theories (see e.g. \cite{Green2012-ul, polchinski_1998, Lust:1989tj})
dominated the research on extra dimensions consisting a solid framework. In
particular the heterotic string theory \cite{GROSS1985253} (defined in ten dimensions) was the
most promising, since potentially it admits experimental compatibility, due to
the fact that the Standard Model (SM) gauge group can be accommodated into those of GUTs that emerge after the dimensional reduction of the initial
$E_8 \times E_8$. It is worth noting that even before the formulation of superstring
theories, an alternative framework was developed that focused on the dimensional
reduction of higher-dimensional gauge theories. This provided another venue for
exploring the unification of fundamental interactions \cite{forgacs,KAPETANAKIS19924,Kubyshin:1989vd,SCHERK197961,MANTON1981502,CHAPLINE1982461,LUST1985309,Manousselis_2004,Chatzistavrakidis:2009mh,Irges:2011de}. The endeavor to
unify fundamental interactions, which shared common objectives with the
superstring theories, was first investigated by Forgacs-Manton (F-M) and
Scherk-Schwartz (S-S). F-M explored the concept of Coset Space Dimensional
Reduction (CSDR) \cite{forgacs}, which can lead naturally to chiral fermions while S-S
focused on the group manifold reduction \cite{SCHERK197961}, which does not admit chiral
fermions. Recent attempts towards realistic models that can be confronted with
experiment can be found in \cite{Irges:2011de,Manolakos:2020cco,Patellis:2023npy}.
\par
On the gravity side diffeomorphism-invariant gravity theory is obviously
invariant with respect to transformations whose parameters are functions of
spacetime, just as in the local gauge theories. Then, naturally, it has been
long believed that general relativity (GR) can be formulated as a gauge theory \cite{PhysRev.101.1597, Kibble:1961ba, PhysRevLett.38.739} with the spin connection as the corresponding gauge field which would
enter in the action through the corresponding field strength. This idea was used
heavily in supergravity (see e.g. \cite{freedman_vanproeyen_2012}) while recently it was employed in
non-commutative gravity too \cite{Chatzistavrakidis_2018, Manolakos:2019fle, Manolakos:2021rcl}.
Along the same lines rather recently was suggested a new idea for unification of
all known interactions in four dimensions. Usually the dimension of the tangent
space is taken to be equal to the dimension of the curved manifold. However, the tangent group of a manifold of dimension $d$ is not necessarily $SO_d$ \cite{Weinberg:1984ke}. It is possible to embed the coordinate tangent space in a higher-dimensional space, and therefore promote the gauge symmetry to a higher isometry group. In refs \cite{Percacci:1984ai, Percacci_1991, Nesti_2008, Nesti_2010, Krasnov:2017epi, Chamseddine2010, Chamseddine2016}, the authors have considered
higher-dimensional tangent spaces in $4-$dimensional spacetime and managed in
this way to achieve unification of internal interactions with gravity. The
geometric
unification of gravity and gauge internal interactions in refs \cite{Chamseddine2010,Chamseddine2016} is realized by writing
the action of the full theory in terms only of the curvature invariants of the
tangent group, which contain the Yang-Mills actions corresponding to the gauge
groups describing in this way together the GR and the internal GUT in a unified
manner. The best model found so far that unifies gravity and a chiral GUT is
based on $SO_{(1,13)}$ in a 14-dimensional
tangent space leading to unification of gravity with $SO_{10}$. However as a
drawback was considered the fact that fermions appear in double representations
of the spinor $16$ of $SO_{10}$, which only means that fermions appear in even
families though \cite{Manolakos:2023hif}. Trying to resolve this problem by imposing Majorana
condition in addition to Weyl in ref \cite{Nesti_2008, Nesti_2010} was proposed instead as a unifying
group the $SO_{(3,11)}$, which leads to the unavoidable appearance of ghosts due to
the more than one time-like coordinates of the Lorentz group.
  Here instead we propose as a unifying group the $SO_{(1,17)}$, in which one can
impose both Weyl and Majorana conditions and the final group obtained in four dimensions is the ordinary $SO_{10}$ GUT \cite{FRITZSCH1975193}, followed by a global $SU_2\times SU_2$ symmetry.

\section{The \texorpdfstring{$SO_{(1,17)}$}{so17} as unifying group}

\subsection{Geometrical construction}\label{a}
Starting with $SO_{(1,17)}$ as the initial gauge symmetry group, we wish to produce symmetry breakings that will lead to the product of two symmetries, one describing gravity as a gauge theory, and the other describing the internal interactions. These breakings can occur via a SSB mechanism, or by imposing constraints to the theory. In order for the presentation of the model to be self-contained and amplify the latter (let's call it "soldering mechanism") over the well-known SSB mechanism, here we lay out and follow the analysis of \cite{Chamseddine2010,Chamseddine2016}, implemented for a $18-$dimensional extended tangent space. The breakings of the present model via the usual Higgs mechanism is also described in the subsection \ref{d}.

At every point of a curved $4-$dimensional Lorenzian metric space, we erect an $18-$dimensional extended tangent space, following a construction analogous to 
\cite{Chamseddine2010,Chamseddine2016}.
The extended tangent space is spanned by the vectors $\mathbf{v}_A$, where $A=0,...,17$ in such a way that the coordinate tangent space, spanned by the coordinate vectors $e_\mu \equiv {\partial}{/}{\partial x^\mu}$, $\mu=0,...,3$, is fully embedded\footnote{This structure can be formulated more mathematically by the fibre bundle theory.}. 
Having chosen $SO_{(1,17)}$ as the structure group of the extended tangent homomorphisms, the scalar product of the basis vectors $\{\mathbf{v}_A\}$ should be orthonormal with respect to the extended, $18-$dim Minkowskian metric, $\eta_{A B}=\text{diag}(-1,+1,...,+1)$,
\begin{equation}\label{eta}
\mathbf{v}_A \cdot \mathbf{v}_B=\eta_{A B}.
\end{equation}
It is clear that the orthogonality of the basis vectors $\{\mathbf{v}_A\}$ is preserved under the extended $SO_{(1,17)}$ Lorentz transformations.

It will prove convenient to separate the tangent space spanned by the basis vectors $\{\mathbf{v}_A\} $ into two orthogonal subspaces. The first is identified with the coordinate tangent space and spanned by the coordinate basis vectors $\{\mathbf{e}_\mu\}$, which are orthonormal with respect to the metric of the base manifold,
\begin{equation}\label{g}
\mathbf{e}_\mu \cdot \mathbf{e}_\nu=g_{\mu \nu}\left(x\right) .
\end{equation}
The second, that will be called internal tangent space, will be the orthogonal complement to the first and spanned by the set of 14 basis vectors $\{\mathbf{n}_{i}\}$, where $i=4,...,17$, which are orthonormal with respect to the Euclidean metric,
\begin{equation}\label{delta}
\mathbf{n}_{i} \cdot
\mathbf{n}_{j}
=\delta_{ij}.
\end{equation}
The projections of the extended tangent space basis vectors, $\{\mathbf{v}_A\}$, onto the embedded coordinate tangent space basis vectors $\{e_\mu\}$, are performed via the soldering forms $\tensor{e}{^A_\mu}$,
\begin{equation}
\mathbf{e}_\mu=
\mathbf{v}_A\tensor{e}{^A_\mu}.
\end{equation}
Then, with the aid of \eqref{eta} one obtains
\begin{equation}\label{sf}
e_{A \mu}=\mathbf{v}_A \cdot \mathbf{e}_\mu.
\end{equation}
Multiplying both sides of \eqref{sf} with $g^{\mu\nu}$ yields,
\begin{equation}
\tensor{e}{_A^\mu}=\mathbf{v}_A\cdot\mathbf{e}^\mu,
\end{equation}
where the $\mathbf{v}_A$ are now projected onto the co-tangent basis, $\{e^\mu\}\equiv \{dx^\mu\}$. Naturally, the soldering forms for which both indices are coordinate, we have, $\tensor{e}{_\nu ^\mu}=\tensor{\delta}{_\nu^\mu}$.
 
The projections of the extended tangent space basis vectors $\{\mathbf{v}_A\}$, onto the basis vectors $\{\mathbf{n}_{i}\}$ are performed via $\tensor{n}{^A_i}$,
\begin{equation}
\mathbf{n}_{i}=
\mathbf{v}_A\tensor{n}{^A_i},
\end{equation}
which with the aid of \eqref{delta} lead to
\begin{equation}\label{sf2}
n_{Ai}=\mathbf{v}_A \cdot \mathbf{n}_{i}.
\end{equation}
By multiplying both sides of \eqref{sf2} with $\delta^{ij}$ one obtains,
\begin{equation}
\tensor{n}{_A^i}=\mathbf{v}_A\cdot\mathbf{n}^i.
\end{equation}
Hence, a vector $\mathbf{v}_A$ can be decomposed into the basis vectors $\mathbf{e}_\mu$ and $\mathbf{n}_{i}$ as
\begin{equation}\label{v}
\mathbf{v}_A=\tensor{e}{_A^\mu}
\mathbf{e}_\mu + 
\tensor{n}{_A^i}\mathbf{n}_{i}.
\end{equation}
Using \eqref{g} and \eqref{delta} 
we obtain expressions for the base manifold metric exclusively in terms of the soldering forms $\tensor{e}{^A_\mu}$,
\begin{equation}\label{metric-vielbeins}
g_{\mu \nu}=\eta_{AB}\tensor{e}{^A_\mu}\tensor{e}{^B_\nu}=\tensor{e}{^A_\mu}
\tensor{e}{_{A_\nu}},
\end{equation}
and, respectively, for the Euclidean metric exclusively in terms of the forms $\tensor{n}{^A_i}$,
\begin{equation}\label{metric-vielbeins2}
\delta_{ij}=\eta_{AB}
\tensor{n}{^A_i}\tensor{n}{^B_j}=
\tensor{n}{^A_i}
\tensor{n}{_{Aj}}.
\end{equation}
As remarked in \cite{Chamseddine2010}, attention should be paid to the fact that  $\tensor{e}{_A^\mu}$ is not inverse to $\tensor{e}{^A_\mu}$ when the dimensions
of the tangent space and the base manifold don't match\footnote{Recalling the definition of $\tensor{e}{^A_\mu}$ as a projector of 18-dimensional vectors onto the 4-dimensional tangent space, it is clear that $\tensor{e}{_A^\mu}$ cannot be considered as a reversed projector operator, since a lower dimensional vector cannot be projected onto a space of higher dimensionality. As already stated, $\tensor{e}{_A^\mu}$ are projections of the vectors $\mathbf{v}_A$ onto the co-tangent space $\{e^a\}$.}. 
In other words, although it is obvious from \eqref{metric-vielbeins} that
\begin{equation}
\tensor{e}{_A^\mu}\tensor{e}{^A_\nu}
=\delta_\nu^\mu,
\end{equation}
when contracting with respect to the tangent indices, it is also clear from \eqref{eta} and \eqref{v}, that
\begin{equation}\label{econtraction}
\tensor{e}{_A^\mu}\tensor{e}{^B_\mu}
=\delta_B^A - 
\tensor{n}{_A^j}\tensor{n}{^B_j},
\end{equation}
given the orthonormality relations
\begin{equation}
n_{j}^A e_A^\mu=0, \quad n_{j}^A n_A^{i}=\delta_{j}^{i}. 
\end{equation}

Parallel transport is defined via the action of affine and spin connections, on the coordinate and extended tangent space basis vectors respectively,
\begin{equation}\label{Parallel Transport}
\nabla_\nu \mathbf{e}_\mu=
\tensor{\Gamma}{^\lambda_{\nu\mu}}
 \mathbf{e}_\lambda, \quad \nabla_\nu \mathbf{v}_A=-
 \tensor{\omega}{_{\nu A}^B}
 \mathbf{v}_B,
\end{equation}
where $\nabla_{\nu}$ is the covariant derivative along the direction of the tangent basis vector ${\mathbf{e}_\nu}$. 

By defining the parallel transport of the coordinate basis vectors as above, a constraint has actually been imposed to the geometrical construction, as the most general form of it would be $\nabla_\nu \mathbf{e}_\mu=
\tensor{\Gamma}{^\lambda_{\nu\mu}}
 \mathbf{e}_\lambda + \tensor{B}{^i_{\nu\mu}}\mathbf{n}_i$. The imposed constraint, $\tensor{B}{^i_{\nu\mu}}=0$, causes the breaking
\begin{equation}
SO_{(1,17)}\rightarrow SO_{(1,3)}\times SO_{14}.
\end{equation}

\par
Having imposed this constraint, can be shown \cite{Chamseddine2016} that the covariant derivative of the internal basis vectors is also an element of the internal subspace. Hence we have as well,  
\begin{equation}\label{nablaa}
\nabla_\nu \mathbf{n}_{i}=-
\tensor{A}{_{\nu i}^{j}}\mathbf{n}_{j}.
\end{equation}
The covariant derivative when acting on scalars, naturally coincides with the ordinary derivative\footnote{In the current analysis, tensor components are considered scalar functions. See also \cite{Nakahara}.}. Therefore, the metricity condition,
\begin{equation}\label{metricity condition}
\nabla_{\nu} \eta_{AB}=0,
\end{equation}
 must hold. Following \eqref{Parallel Transport}, the action of the covariant derivative operator on a tangent vector
$V=V^\mu\mathbf{e}_\mu$
is, 
\begin{eqnarray}
\nabla_\nu\left(V^\mu\mathbf{e}_\mu\right)=
\left(\partial_\nu V^\mu\right)\mathbf{e}_\mu+V^\mu\nabla_\nu\mathbf{e}_\mu=
\left(
\partial_\nu V^\mu+
V^\lambda\tensor{\Gamma}{^{\mu}_{\nu\lambda}}
\right)\mathbf{e}_\mu=
\left(\nabla_\nu V\right)^\mu\mathbf{e}_\mu.
\end{eqnarray}
It is convenient to use a different symbol to express the total action of the covariant derivative on a vector as an action upon its components\footnote{In general, the $\nabla$ operator performs parallel transport in a passive way, acting on the basis vectors as denoted in \eqref{Parallel Transport} and 
\eqref{nablaa}, treating the tensor components as scalar functions. In contrast, the $D$ operator performs parallel transport in an active way, acting as a covariant derivative upon the tensor components leaving the basis vectors intact, i.e. \eqref{VectorCD}, \eqref{CovectorCD}.}. Guided from the above equation we define, 
\begin{equation}\label{VectorCD}
D_\nu V^\mu\equiv\left(\nabla_\nu V\right)^\mu=
\partial_\nu V^\mu+
\tensor{\Gamma}{^{\mu}_{\nu\lambda}}V^\lambda .
\end{equation}
Similarly, for co-vectors $W=W_\mu dx^\mu$, we should have, 
\begin{equation}\label{CovectorCD}
D_\nu W_\mu=
\partial_\nu W_\mu-
W_\lambda\tensor{\Gamma}{^{\lambda}_{\nu\mu}}.
\end{equation}
Analogous equations hold for the action of the covariant derivative on the components of an extended tangent space spinor, $\psi=\psi^A\mathbf{v}_A$, now with the 
spin connection performing parallel transportation,
\begin{equation}\label{SpinorCD}
D_\nu\psi^A=\partial_\nu\psi^A-\psi^B
\tensor{\omega}{_{\nu B}^A},
\end{equation}
and for the the action of the covariant derivative on the components of coordinate space spinors,
\begin{equation}\label{SpinorVCD}
D_\nu X^{i}=\partial_\nu X^{i}-X^{j}
\tensor{A}{_{\nu j}^{i}}.
\end{equation}

Simultaneous validity of  \eqref{eta} and \eqref{metricity condition}, implies that the spin connection is antisymmetric under the interchange of its extended tangent space indices,
\begin{equation}\label{spin connection antisymmetry}
\omega_{\nu A B}=-\omega_{\nu B A}.
\end{equation}
Covariant differentiation of \eqref{sf} and taking into account \eqref{Parallel Transport} implies,
\begin{equation}\label{partiale}
\partial_\nu e_{A \mu}=-\tensor{\omega}{_{\nu A}^B} e_{B \mu}+
\tensor{\Gamma}{^\lambda_{\mu\nu}}e_{A \lambda} .
\end{equation}
In perfect analogy, covariant differentiation of \eqref{sf2} yields the equation
\begin{equation}\label{partialn}
\partial_{\mu} n_{Ai}=-\omega_{\mu A}{ }^{B} n_{Bi}-A_{\mu i}{}^{j}n_{Aj}.
\end{equation}

Equations \eqref{partiale} are $4\times 18\times 4=288$ in number and should be solved for the affine and spin connection components, in terms of the given soldering forms. 
In a torsion-free base manifold, the affine connection is symmetric over the interchange of its lower indices,
\begin{equation}\label{torsionless base manifold}
\tensor{\Gamma}{^\lambda_{\mu\nu}}=
\tensor{\Gamma}{^\lambda_{\nu\mu}}.
\end{equation}
This means that the number of independent affine connection components are $4\times 10=40$  and can be determined separately. 

Let us first operate with the covariant derivative $\nabla_\lambda$ on \eqref{metric-vielbeins},
\begin{equation}\label{nablag}
\nabla_\lambda g_{\mu \nu}=\partial_\lambda g_{\mu \nu}=\partial_\lambda\left(
\tensor{e}{^A_\mu}e_{A \nu}\right)=\left(\partial_\lambda\tensor{e}{^A_\mu}\right)e_{A \nu}+\tensor{e}{^A_\mu}\partial_\lambda e_{A \nu}.
\end{equation}
Then, combining \eqref{partiale} and \eqref{nablag}, we obtain 
\begin{equation}
\tensor{\Gamma}{^\rho_{\mu\lambda}}
\tensor{g}{_{\rho\nu}}+
\tensor{\Gamma}{^\rho_{\nu\lambda}}g_{\mu\rho}
=\partial_\lambda g_{\mu\nu},
\end{equation}
which when inverted, gives the explicit expression for the torsion-free Christofell connection in terms of the metric, \begin{equation}\label{christofell}
\tensor{\Gamma}{^\gamma_{\mu\nu}}=
\frac{1}{2} g^{\lambda \rho}\left(g_{\mu\rho, \nu}+g_{\rho\nu,\mu}-g_{\mu \nu,\rho}\right).
\end{equation}
What we have actually done here was to employ 40 out of the 288 equations \eqref{partiale} to fully determine $\tensor{\Gamma}{^\gamma_{\mu\nu}}$.
Therefore, we are now left with $248$ equations to determine the components of the spin connection which is antisymmetric over its last two indices \eqref{spin connection antisymmetry}. Since the
$\nu$ index runs over the 4 base manifold dimensions, and the indices $A,B$ over the 18 extended tangent space dimensions, we see that there are $612$ components to be determined, while there are only $248$ equations in our disposal. This implies that we are left with $612-248=364$ undefined spin connection components, a number which matches the $SO_{14}$ gauge fields (multiplied by $4$).

The above results confirm that the initial group, $SO_{(1,17)}$, is defined to have an inner structure related to the geometry of the coordinate manifold. Its first 4 dimensions correspond to the tangent space of the manifold, while the rest $14$ remain unmixed with them, showing that the initial gauge group has been reduced to the direct product $SO_{(1,3)}\times SO_{14}$. By the first group, $SO_{(1,3)}$, we are going to describe the spacetime geometry, while by $SO_{14}$ the internal interactions.

The local Lorentz transformation law of a basis vector, $\mathbf{v}_A$, is
\begin{equation}\label{lorentzv}
\mathbf{v}_A \rightarrow \tilde{\mathbf{v}}_A=\tensor{\Lambda}{_A^B} \mathbf{v}_B,
\end{equation}
 thus the soldering forms transform covariantly, as 
\begin{equation}
\tensor{e}{_A^\mu}
 \rightarrow
\tensor{\tilde{e}}{_A^\mu}=\tensor{\Lambda}{_A^B} \tensor{e}{_B^\mu}.
\end{equation}
Using the local Lorentz transformation, \eqref{lorentzv}, and the parallel transportation rule, \eqref{Parallel Transport}, of the basis vectors, we can also show that 
\begin{equation}
\omega_{\nu A}{ }^B \rightarrow \tilde{\omega}_{\nu A}{ }^B=\left(\Lambda \omega_\nu \Lambda^{-1}\right)_A{ }^B+\left(\Lambda \partial_\nu \Lambda^{-1}\right)_{A}{ }^B ,
\end{equation}
which explicitly shows that the spin connection transforms under local Lorentz transformations as a Yang-Mills field, with $SO_{(1,17)}$ 
in the role of the gauge group.

\subsection{Constructing the total action}
Taking into account the expressions \eqref{VectorCD} and \eqref{CovectorCD},
we can derive the explicit form for the affine curvature tensor, following the definition,
\begin{equation}
[D_\nu,D_\lambda]V^\mu=V^\rho\tensor{R}{^\mu_{\rho\nu\lambda}}.
\end{equation}
Indeed, after some straightforward algebra we obtain, 
\begin{equation}
[D_\nu,D_\lambda]V^\mu=V^\rho\left(
\partial_\nu\tensor{\Gamma}{^\mu_{\rho\lambda}}-
\partial_\lambda\tensor{\Gamma}{^\mu_{\rho\nu}}+
\tensor{\Gamma}{^\sigma_{\rho\lambda}}
\tensor{\Gamma}{^\mu_{\sigma\nu}}-
\tensor{\Gamma}{^\sigma_{\rho\nu}}
\tensor{\Gamma}{^\mu_{\sigma\lambda}}
\right),
\end{equation}
thus, it is clear that we can identify the affine curvature tensor, 
\begin{equation}
\tensor{R}{^\mu_{\rho\nu\lambda}}=
\partial_\nu\tensor{\Gamma}{^\mu_{\lambda\rho}}-
\partial_\lambda\tensor{\Gamma}{^\mu_{\nu\rho}}+
\tensor{\Gamma}{^\mu_{\nu\sigma}}
\tensor{\Gamma}{^\sigma_{\lambda\rho}}-
\tensor{\Gamma}{^\mu_{\lambda\sigma}}
\tensor{\Gamma}{^\sigma_{\nu\rho}}.
\end{equation}
Similarly, following the definition for the spin curvature tensor, 
\begin{equation}\label{ddpsi}
[D_\mu,D_\nu]\psi^A=
\psi^B\tensor{R}{_{\mu\nu}^A_B},
\end{equation}
and employing \eqref{SpinorCD} we obtain, 
\begin{equation}\label{Spinor Curvature}
\tensor{R}{_{\mu\nu}^{AB}}=
\partial_\mu\tensor{\omega}{_\nu^{AB}}-\partial_\nu\tensor{\omega}{_\mu^{AB}}+
\tensor{\omega}{_\mu^A_C}\tensor{\omega}{_\nu^{CB}}-
\tensor{\omega}{_\nu^A_C}\tensor{\omega}{_\mu^{CB}}.
\end{equation}
Finally, following the definition for the coordinate curvature tensor,
\begin{equation}\label{ddx}
[D_\mu,D_\nu]X^{i}=
X^{j}\tensor{F}{_{\mu\nu}^{i}_{j}},
\end{equation}
and employing \eqref{SpinorVCD} we get, 
\begin{equation}\label{Vertical Curvature}
\tensor{F}{_{\mu\nu}^{ij}}=
\partial_\mu\tensor{A}{_\nu^{ij}}-\partial_\nu\tensor{A}{_\mu^{ij}}+
\tensor{A}{_\mu^{i}_{k}}
\tensor{A}{_\nu^{kj}}-
\tensor{A}{_\nu^{i}_{k}}
\tensor{A}{_\mu^{kj}}.
\end{equation}

Taking the partial derivative $\partial_\rho$ of \eqref{partiale} and substracting the same result but with the indices $\nu$ and $\rho$ interchanged, we
end up to a relation between the spin and affine curvature tensors,
\begin{equation}\label{spin-affine curvatures}
\tensor{R}{_{\mu\nu}^{AB}}(\omega) e_{B \lambda}=\tensor{e}{^A_\rho}
\tensor{R}{_{\mu\nu}^{\rho}_{\lambda}}(\Gamma) .
\end{equation}
Following analogous procedure in \eqref{partialn} we end up to a relation among the spin and coordinate curvature tensors,
\begin{equation}\label{spin-vertical curvatures}
  \tensor{n}{_B^i}\tensor{R}{_{\mu\nu A}^B}(\omega)= \tensor{n}{_A^j}\tensor{F}{_{\mu\nu j}^i}(A).
\end{equation}
Employing \eqref{econtraction}, equation \eqref{spin-affine curvatures} can be rewritten as an expansion on the basis vectors of the two subspaces of the extended tangent space,
\begin{equation}\label{rrr}
\tensor{R}{_{\mu\nu}^{AB}}(\omega)=\tensor{R}{_{\mu \nu}^{A C}}(\omega)\tensor{n}{_C^i}\tensor{n}{^B_i}+\tensor{R}{^{\rho}_{\lambda\mu\nu}}(\Gamma) \tensor{e}{^A_\rho} e^{B \lambda} .
\end{equation}
Substituting \eqref{spin-vertical curvatures} in the equation above, we obtain
\begin{equation}
\tensor{R}{_{\mu\nu}^{AB}}(\omega)=
\tensor{n}{^A_i}\tensor{n}{^B_j}
\tensor{F}{_{\mu\nu}^{ij}}(A)+
\tensor{e}{^A_\rho}
\tensor{e}{^{B\lambda}}
\tensor{R}{_{\mu\nu}^{\rho}_{\lambda}}(\Gamma).
\end{equation}
We see that the spin curvature has been completely decomposed into the coordinate and affine curvatures. 
From this expression, all the invariants of the theory up to second order will be produced.

For the first order, the only contraction possible is 
\begin{equation}
\tensor{R}{_{\mu\nu}^{AB}}(\omega)\tensor{e}{_A^\mu}\tensor{e}{_B^\nu}=R(\Gamma),
\end{equation}
which produces the Ricci scalar of the theory. The second order invariants are 
\begin{equation}
R^2(\Gamma),\; R_{\mu \nu}(\Gamma) R^{\mu \nu}(\Gamma),\; R_{\mu \nu \lambda \delta}(\Gamma)R^{\mu \nu \lambda \delta}(\Gamma),
\end{equation}
which get produced by various combinations of soldering forms acting on the curvature 2-form. The kinetic terms are going to be produced by the contraction
\begin{equation}
g^{\mu \lambda} g^{\nu \delta} R_{\mu \nu}{ }^{A B}(\omega) R_{\lambda \delta A B}(\omega)=g^{\mu \lambda} g^{\nu \delta}\left(F_{\mu \nu}{}^{i j}(A) F_{\lambda \delta i j}(A)\right)+R_{\mu \nu \lambda \delta}(\Gamma) R^{\mu \nu \lambda \delta}(\Gamma).
\end{equation}
Now we have produced all the curvature invariants up to second order. In the general action, $512-$dimensional Dirac spinor fields also have to be included,
\begin{equation}
\int d^4 x \sqrt{-g} \bar{\psi} i \Gamma^A \tensor{e}{_A^\mu}D_\mu \psi,
\end{equation}
where $\Gamma_A$ matrices satisfy the Clifford algebra
\begin{equation}
\left\{\Gamma^A, \Gamma^B\right\}=2 \eta^{A B},
\end{equation}
and
\begin{equation}
D_\mu \equiv \partial_\mu+\frac{1}{4} 
\tensor{\omega}{_\mu^{AB}}
S_{AB},
\end{equation}
where $S_{AB}=\frac{1}{2}[\Gamma_A,\Gamma_B]$ are the generators of the $SO_{(1,17)}$ algebra.

Hence, the expression of the general action of the theory is 

\begin{align} & I_{SO_{(1,17)}}=\int d^4 x \sqrt{-g}\left[\frac{1}{16 \pi G} R_{\mu \nu}{ }^{A B}(\omega) 
\tensor{e}{_A^\mu}
\tensor{e}{_B^\nu}
\right. \nonumber \\ 
& \quad+R_{\mu \nu}{ }^{A B} R_{\lambda \delta}{ }^{C D}\left(a\;
\tensor{e}{_A^\mu}
\tensor{e}{_B^\nu}
\tensor{e}{_C^\lambda}
\tensor{e}{_D^\delta}
+b\;
\tensor{e}{_A^\mu}
\tensor{e}{_C^\nu}
\tensor{e}{_B^\lambda}
\tensor{e}{_D^\delta}
+c\;
\tensor{e}{_C^\mu}
\tensor{e}{_D^\nu}
\tensor{e}{_A^\lambda}
\tensor{e}{_B^\delta}
\right) \\ 
& \left.\quad-\frac{1}{4} g^{\mu \lambda} g^{\nu \delta} R_{\mu \nu}{ }^{A B}(\omega) R_{\lambda \delta A B}(\omega) +\bar{\psi} i \Gamma^A \tensor{e}{_A^\mu} D_\mu \psi\right] \Rightarrow \nonumber \\ \nonumber\\ 
 & I_{SO_{(1,3)}\times SO_{14}}=\int d^4 x \sqrt{-g}\bigg[\frac{1}{16 \pi G} R(\Gamma) +a R^2(\Gamma) -b R_{\mu \nu}(\Gamma) R^{\mu \nu}(\Gamma)+ \nonumber
 \\ & \quad + (c-\frac{1}{4}) R_{\mu \nu \lambda \delta}(\Gamma)R^{\mu \nu \lambda \delta}(\Gamma) -\frac{1}{4}g^{\mu\lambda}g^{\nu \delta}F_{\mu\nu}{}^{ij}(A)F_{\lambda\delta ij}(A)+ \\ & \quad + \bar{\psi}_{SO_{(1,3)}} i {\Gamma}^\mu D_\mu \psi_{SO_{(1,3)}} + \bar{\psi}_{SO_{14}} i {\Gamma}^j \tensor{e}{_j^\mu} D_\mu \psi_{SO_{14}}\bigg], \nonumber
 \end{align}  
where $a,\; b, \; c$ are dimensionless constants, and the Weyl representation has been chosen for the Gamma matrices. The above action consists of $SO_{(1,3)}$ and $SO_{14}$ invariants, as expected. By setting $a=\frac{b}{4}=c-\frac{1}{4}$, the curvature terms form the integrand of the Gauss-Bonnet topological invariant, hence they do not contribute to the field equations, avoiding in this way the appearance of ghosts \cite{Alvarez_Gaume_2016}. By the $SO_{(1,3)}$ part we are able to retrieve Einstein's gravity as a gauge theory \cite{PhysRev.101.1597, Kibble:1961ba, PhysRevLett.38.739}, while by the $SO_{14}$ we are going to describe internal interactions.

\subsection{Breakings}\label{d}

  According to the subsection \ref{a}, the original gauge symmetry, $SO_{(1,17)}$, of the
theory, is being reduced to $SO_{(1,3)} \times SO_{14}$ by employing the
soldering mechanism presented above, i.e. the
$C_{SO_{(1,17)}} (SO_{(1,3)}) = SO_{14}$ remains as
the gauge group that will describe the internal interactions. The same breaking can occur via Higgs mechanism, by introducing a scalar field in the $170$ representation\footnote{The representation $3060$ can be also used for that purpose. The two breakings differ on the presence of a remaining unbroken parity symmetry in the case of $170$, of which the $SO_{10}$ analogous is discussed in detail in \cite{PhysRevLett.52.1072, PhysRevD.31.1718}.} of $SO_{18}$ \cite{Feger_2020}, and with the help of a Lagrange multiplier
we can break the gauge symmetry non-linearly \cite{PhysRevD.21.1466,Manolakos:2021rcl},
\begin{equation}
   \begin{gathered}
    SO_{18} \rightarrow SU_2 \times SU_2 \times SO_{14}\\
170 = (1,1,1) + (3,3,1) +(2,2,14) +(1,1,104).
\end{gathered}
\end{equation}

In order to break further the resulting $SO_{14}$ gauge symmetry to a symmetry
of a more familiar GUT, such as $SO_{10}$, we can employ a second 
Higgs mechanism by using the $104$ representation\footnote{Had we chosen the $3060$ of $SO_{18}$ for the previous breaking, the appropriate representation for the breaking of $SO_{14}$ would be $1001$ with analogous effects on the remaining symmetry.} of $SO_{14}$ \cite{Slansky:1981yr, McKay1981TablesOD},
\begin{equation}
\begin{gathered}
    SO_{14} \rightarrow SU_2 \times SU_2 \times SO_{10}\\
104 = (1, 1, 1) + (3, 3, 1) + (2, 2, 10) + (1, 1, 54).
\end{gathered}
\end{equation}

Of the rest $SU_2 \times SU_2 \times SU_2 \times SU_2$ symmetry that remains, one part $SU_2\times SU_2$ should be used for describing gauge gravity, while the other $SU_2\times SU_2$ should be broken. The irreducible spinor representation of $SO_{18}$ is $256$, which under $SU_2 \times SU_2 \times SO_{14}$, decomposes as
\begin{equation}
    256=(2,1,64)+(1,2,\Bar{64}),
\end{equation}
while the irreducible spinor representation of $SO_{14}$ is $64$, that under $SU_2 \times SU_2 \times SO_{10}$ decomposes as
\begin{equation}
    64=(2,1,16)+(1,2,\Bar{16}).
\end{equation}
 By introducing further two scalars in the $256$
representation of $SO_{18}$, when they take VEVs in their $(\langle 2 \rangle,1,16)$ and
$(1,\langle 2 \rangle,\bar{16})$ components of $SO_{14}$ under the $SU_2 \times SU_2 \times SO_{10}$ decomposition, the final unbroken gauge symmetry is $SO_{10}$. The final total symmetry that we are left with is
\begin{equation}
    [SU_2 \times SU_2]_{\text{Lorentz}}\times [SU_2 \times SU_2]_{\text{Global}} \times SO_{10}{}_{\text{Gauge group}}.
\end{equation}

\section{Weyl and Majorana spinors}
A Dirac spinor, $\psi$, has $2^{D/2}$ independent components in $D$ dimensions. The Weyl and Majorana constraints each divide the number of independent components by $2$.
The Weyl condition can be imposed only for $D$ even, so a Weyl-Majorana spinor
has $2^{(D-4)/2}$ independent components (when $D$ is even). Weyl-Majorana spinors can exist only for $D = 4n+2$; real Weyl-Majorana spinors
can exist for $D = 2$ modulo $8$, and pseudoreal Weyl-Majorana spinors can exist
for $D = 6$ modulo $8$.

The unitary representations of the Lorentz group $SO_{(1, D-1)}$ are labeled by a
continuous momentum vector $\mathbf{k}$, and by a spin "projection", which in $D$ dimensions
is a representation of the compact subgroup $SO_{(D-2)}$.  The Dirac, Weyl,
Majorana, and Weyl-Majorana spinors carry indices that transform as
finite-dimensional non-unitary spinor representations of $SO_{(1, D-1)}$.

It is well-known (see e.g. \cite{D_Auria_2001} for a review, or \cite{majoranaspinors} for a more concise
description) that the type of spinors one obtains for $SO_{(p, q)}$ in the real
case is governed by the signature $(p-q)$ mod $8$. Among even signatures,
signature $0$ gives a real representation, signature $4$ a quaternionic
representation, while signatures $2$ and $6$ give complex representations. In the case of $SO_{(1,17)}$ the signature is zero, and the imposition of the Majorana condition on the spinors is permitted, in addition to Weyl.

Let us recall for completeness and fixing the notation, the otherwise well-known case of $4$ dimensions. The $SO_{(1,3)}$ spinors in the usual $SU_{2}\times SU_{2}$
basis transform as $(2,1)$ and $(1,2)$, with representations labeled by their
dimensionality. The $2-$component Weyl spinors, $\psi_L$ and $\psi_R$, transform as the
irreducible spinors,
\begin{equation}
\psi_L \thicksim (2, 1), \quad \psi_R \thicksim (1, 2),       
\end{equation}
of $SU_{2}\times SU_{2}$ with "$\thicksim$" meaning "transforms as".
A Dirac spinor, $\psi$, can be made from the direct sum of $\psi_L$ and $\psi_R$,
\begin{equation}
  \psi \thicksim (2,1) \oplus (1,2).     
  \end{equation}

In $4$-component notation the Weyl spinors in the Weyl basis are $(\psi_L, 0)$ and $(0,\psi_R)$, and are eigenfunctions of $\gamma^5$ with eigenvalues $+ 1$ and $- 1$, respectively.

The usual Majorana condition for a Dirac spinor has the form,
\begin{equation}\label{3}
  \psi = C {\bar{\psi}}^{T},   
  \end{equation}
where $C$ is the charge-conjugation matrix.
In $4$ dimensions $C$ is off-diagonal in the Weyl basis, since it maps the
components transforming as $(2,1)$ into $(1,2)$. Therefore, if one tries to
impose \eqref{3} on a Weyl spinor, there is no non-trivial solution and
therefore Weyl-Majorana spinors do not exist in $4$ dimensions.

  For $D$ even, it is always possible to define a Weyl basis where $\Gamma^{D+1}$
(the product of all $D$ $\Gamma$ matrices) is diagonal, so
\begin{equation}\label{4}
  \Gamma^{D+1} \psi_\pm = \pm \psi_\pm.  \end{equation}

We can express $\Gamma^{D+1}$ in terms of the chirality operators in 4 and extra $d$
dimensions,
\begin{equation}\label{5}
  \Gamma^{D+1} = \gamma^5 \otimes \gamma^{d+1}.      
  \end{equation}

Therefore the eigenvalues of $\gamma^5$ and $\gamma^{d+1}$ are interrelated. However, clearly
the choice of the eigenvalue of $\Gamma^{D+1}$ does not impose the eigenvalues on the
interrelated $\gamma^5$ and $\gamma^{d+1}$.

Since $\Gamma^{D+1}$ commutes with the Lorentz generators, then each of the $\psi_+$ and $\psi_-$
transforms as an irreducible spinor of $SO_{(1, D-1)}$. For $D$ even, the $SO_{(1, D-1)}$ always has two independent irreducible spinors; for $D = 4n$ there are two
self-conjugate spinors $\sigma_D$ and 
${\sigma_D}^\prime$, while for $D = 4n + 2$, $\sigma_D$ is
non-self-conjugate and $\bar{\sigma}_D$ is the other spinor. By convention is selected
$\psi_+\thicksim \sigma_D$ and $\psi_-\thicksim {\sigma_D}^\prime$ or $\bar{\sigma}_D$. Accordingly, Dirac spinors are defined as
direct sum of Weyl spinors,

\begin{equation}\label{6}
    \psi  = \psi_+ \oplus \psi_- \thicksim 
    \begin{cases} \sigma_D \oplus {\sigma_D}^\prime \; &\text{for}\; D = 4n \\
 \sigma_D \oplus \bar{\sigma_D}\; &\text{for}\; D = 4n+2.
    \end{cases}         
\end{equation}

When $D$ is odd there are no Weyl spinors, as already mentioned.

The Majorana condition can be imposed in $D = 2,\; 3,\; 4 +8n$ dimensions and
therefore the Majorana and Weyl conditions are compatible only in $D = 4n +
2$ dimensions.

Let us limit ourselves here in the case that $D = 4n+2$, while for the rest one can consult refs \cite{CHAPLINE1982461,KAPETANAKIS19924}. Starting with Weyl-Majorana spinors in
$D = 4n + 2$ dimensions, we are actually forcing a representation $f_R$
of a gauge group defined in higher dimensions to be the charge conjugate of
$f_L$, and we arrive in this way to a $4$-dimensional theory with the fermions only
in the $f_L$ representation of the gauge group.

  In our case, we have for the Weyl spinor of $SO_{(1,17)}$:
  \begin{equation*}
    SO_{(1,17)} \rightarrow [SU_2 \times SU_2 ]_{\text{Lorentz}} \times SO_{14} {}_{\text{Gauge group}}
  \end{equation*}
  
\begin{equation}
   \sigma_{18} = 256 = (2,1; 64) + (1,2; \bar{64}).
\end{equation}

Then, the Majorana condition maps the $(2,1; \bar{64})$ into the $(1,2; 64)$.
Therefore in $4$ dimensions, only the $(2,1; 64)$ remains from the spinor $256$ of
$SO_{18}$, after imposing the Majorana condition, i.e. we obtain $SO_{14}$ with $64_L$.

On the other hand, the spinor of $SO_{14}$, $64_L$, has the following decomposition after the SSB of $SU_2 \times SU_2$, as described in \ref{d} under $[SU_{2} \times SU_{2}]_{\text{Global}} \times SO_{10}$:
\begin{equation}
SO_{14} \rightarrow [SU_{2} \times SU_{2}]_{\text{Global}} \times SO_{10}{}_{\text{Gauge}}    
\end{equation}
\begin{equation}
    64  = (2,1; 16) + (1,2; \bar{16}).
\end{equation}

Therefore, after imposing Weyl (by choosing $\sigma_{18}$) and Majorana conditions in $4$ dimensions, we obtain $f_L = (2,1,16)_L$ and ${(1,2,\bar{16})}_L = {(1,2, 16)}_R =
g_R$. The $f_L$, $g_R$ are eigenfunctions of the $\gamma^5$ matrix with eigenvalues $+1$ and $-1$
respectively, as already mentioned. Keeping only the $+1$ eigenvalue, i.e.
imposing an additional discrete symmetry, we are left with ${(2,1,16)}_L$.

\section{Conclusions}
In the present work we have constructed a realistic model based on the idea that unification of gravity and internal interactions in four dimensions can be achieved by gauging an enlarged tangent Lorentz group. The enlarged group used in our construction is originally the $SO_{(1,17)}$, which eventually, in the broken phase, leads to GR and the $SO_{10}$ GUT accompanied by an $SU_2 \times SU_2$ global symmetry. The latter leads to even number of families.

In the phenomenological analysis that will be presented in a future work, obviously we will include appropriate scalar fields that will (i) break the $SO_{10}$ to the SM and (ii) make the fourth generation of fermions heavy 
in the minimal setting of fermions in the model.

The unifying group $SO_{(1,17)}$ does not contain the whole Poincar\'e group, but rather the Lorentz rotations, $SO_{(1,3)}$. Therefore, a conflict with the Coleman-Mandula (C-M) theorem, which states that internal and spacetime symmetries cannot be mixed \cite{Coleman1967}, is avoided. Recall that the C-M theorem has several hypotheses with the most
relevant being that the theory is Poincar\'e invariant. It might be a challenge of a further study to start with a unifying group that includes translational symmetry and examine consistency with the C-M theorem. 

Another point concerning the phenomenological analysis is that in the present scheme we will do a RG analysis, in which, in addition to
considering the various spontaneous symmetry breakings of $SO_{10}$, the fact that the unification scale is the Planck scale should also be 
taken into account. Finally, we note that the use of the RG analysis is legitimate based on the theorem \cite{Polchinski:1984uw} stating
that if an effective $4-$dimensional theory is renormalizable by power counting, then
it is consistent to consider it as renormalizable a la Wilson \cite{Wilson:1971bg,Wilson:1973jj,Kadanoff:1966wm}.
\subsubsection*{Acknowledgments}
It is a pleasure to thank Pantelis Manousselis for continuous discussions and suggestions on the content of the paper. We would like also to thank Ali Chamseddine, Slava Mukhanov and Roberto Percacci for several discussions on their work. Finally, we would like to thank Peter Forgacs, Dieter Lust and Roberto Percacci for reading the manuscript and for their comments. D.R. would like to thank NTUA for a fellowship for doctoral studies. D.R. and G.Z. have been supported by the Basic Research Programme,
PEVE2020 of National Technical University of Athens, Greece. G.Z. would like to thank CERN-TH, MPP-Munich, ITP-Heidelberg and DFG Exzellenzcluster
2181:STRUCTURES of Heidelberg University for their hospitality and support. 

\printbibliography

@article{Chamseddine2010,
  year = {2010},
  volume = {03},
  author = {A.H. Chamseddine and V. Mukhanov},
  journal = {JHEP},
pages = "033"
}

@article{Chamseddine2016,
  year = {2016},
  volume = "03",
    pages = {020},
  author = {A.H. Chamseddine and V. Mukhanov},
  journal = {JHEP}
}

@book{Nakahara,
author={Nakahara, M.},
 title = {Geometry, Topology and Physics},
publisher={CRC Press},
year={2003}
}

@article{Witten:1983,
    author = "Witten, E.",
    journal = "Conf. Proc. C",
    volume = "8306011",
    year = "1983",
pages = "227"
}

@article{Kaluza:1921,
    author = "Kaluza, Th.",
    journal = "Sitzungsber. Preuss. Akad. Wiss. Berlin (Math. Phys.)",
pages = "966",
    year = "1921"
}

@BOOK{Green2012-ul,
  author    = "Green, M.B. and others",
title = "Superstring Theory",
volume = "1 \& 2",
  publisher = "Cambridge University Press",
  year      =  1988
}

@article{Klein:1926,
    author = "Klein, O.",
    journal = "Z. Phys.",
    volume = "37",
pages = "895",
    year = "1926"
}

@article{Kerner:1968,
    author = "Kerner, R.",
    journal = "Ann. Inst. H. Poincare Phys. Theor.",
    volume = "9",
pages = "143",
    year = "1968"
}

@article{GROSS1985253,
journal = {Nucl. Phys. B},
volume = {256},
pages = "253",
year = {1985},
author = {D.J. Gross and J.A. Harvey and E. Martinec and R. Rohm}
}

@article{CHO1987358,
journal = {J. Math. Phys.},
volume = {16},
year = {1975},
author = {Y.M. Cho},
pages = "2029"
}

@article{PhysRevD.12.1711,
  author = {Cho, Y.M. and Freund, P.G.O.},
  journal = {Phys. Rev. D},
  volume = {12},
  year = {1975},
pages = "1711"
}

@article{PhysRevLett.32.438,
  author = {Georgi, H. and Glashow, S.L.},
  journal = {Phys. Rev. Lett.},
  volume = {32},
  year = {1974},
pages = "438"
}

@article{FRITZSCH1975193,
journal = {Ann. Phys.},
volume = {93},
year = {1975},
author = {H. Fritzsch and P. Minkowski},
pages= "193"
}

@book{polchinski_1998, 
title = "{String theory}",
volume = "{1 \& 2}",
publisher={Cambridge University Press},
author={Polchinski, J.}, year={1998}
}

@article{forgacs,
author = {P. Forg{\aa}cs and N.S. Manton},
volume = {72},
journal = {Commun. Math. Phys.},
pages = "15",
year = {1980},
}

@article{KAPETANAKIS19924,
journal = {Phys. Rept.},
volume = {219},
author = {D. Kapetanakis and G. Zoupanos},
year = {1992},
pages = "4"
}

@book{Kubyshin:1989vd,
    author = "Kubyshin, Y.A. and Volobuev, I.P. and Mourao, J.M. and Rudolph, G.",
title = "Dimensional Reduction of Gauge Theories, Spontaneous Compactification and Model Building",
    publisher = "Springer",
    volume = "349",
    year = "1989"
}

@article{SCHERK197961,
journal = {Nucl. Phys. B},
volume = {153},
year = {1979},
author = {J. Scherk and J.H. Schwarz},
pages = "61"
}

@article{MANTON1981502,
journal = {Nucl. Phys. B},
volume = {193},
year = {1981},
author = {N.S. Manton},
pages = "502"
}

@article{CHAPLINE1982461,
journal = {Nucl. Phys. B},
volume = {209},
year = {1982},
author = {G. Chapline and R. Slansky},
pages = "461"
}

@article{LUST1985309,
journal = {Phys. Lett. B},
volume = {165},
year = {1985},
author = {D. Lust and G. Zoupanos},
pages = "309"
}

@article{Manousselis_2004,
	year = 2004,
	volume = {2004},
	number = {11},
	author = {P. Manousselis and G. Zoupanos},
	journal = {JHEP},
pages = "025"
}

@article{Chatzistavrakidis:2009mh,
    author = "A. Chatzistavrakidis and G. Zoupanos",
    journal = "JHEP",
    volume = "09",
    year = "2009",
pages = "077"
}

@article{Irges:2011de,
    author = "Irges, N. and Zoupanos, G.",
    journal = "Phys. Lett. B",
    volume = "698",
    year = "2011",
pages = "146"
}

@article{Manolakos:2020cco,
    author = "Manolakos, G. and Patellis, G. and Zoupanos, G.",
    journal = "Phys. Lett. B",
    volume = "813",
    year = "2021"
}

@article{Patellis:2023npy,
    author = "Patellis, G. and Porod, W. and Zoupanos, G.",
    eprint = "2307.10014",
    archivePrefix = "arXiv",
    primaryClass = "hep-ph",
    reportNumber = "CERN-TH-2023-124",
    year = "2023"
}

@article{PhysRev.101.1597,
  author = {Utiyama, R.},
  journal = {Phys. Rev.},
  volume = {101},
  year = {1956},
pages = "1597",
  publisher = {American Physical Society},
}

@article{Kibble:1961ba,
    author = "Kibble, T.W.B.",
    journal = "J. Math. Phys.",
    volume = "2",
    year = "1961",
pages = "212"
}

@article{PhysRevLett.38.739,
  author = {MacDowell, S.W. and Mansouri, F.},
  journal = {Phys. Rev. Lett.},
  volume = {38},
pages = "739",
  year = {1977},
  publisher = {American Physical Society},
}

@article{PhysRevD.21.1466,
  author = {Stelle, K.S. and West, P.C.},
  journal = {Phys. Rev. D},
  volume = {21},
  year = {1980},
  pages = {1466},
}

@book{freedman_vanproeyen_2012, 
place = {Cambridge, U.K.},
publisher={Cambridge University Press}, 
author={Freedman, D.Z. and Van Proeyen, A.}, 
title={Supergravity},
year={2012}
}

@article{Chatzistavrakidis_2018,
year = 2018,
volume = {66},
journal = {Fortsch. Phys.},
author = "Chatzistavrakidis, A. and others",
    pages = "1800047"
}

@article{Manolakos:2019fle,
    author = "Manolakos, G. and Manousselis, P. and Zoupanos, G.",
    journal = "JHEP",
    volume = "08",
    year = "2020",
pages = "001"
}

@article{Manolakos:2021rcl,
    author = "Manolakos, G. and Manousselis, P. and Zoupanos, G.",
    journal = "Fortsch. Phys.",
    volume = "69",
    year = "2021",
pages = "2100085"
}

@inproceedings{Weinberg:1984ke,
    author = "Weinberg, S.",
    booktitle = "{Fifth Workshop on Grand Unification}",
    year = "1984"
}

@article{Percacci:1984ai,
    author = "Percacci, R.",
    journal = "Phys. Lett. B",
    volume = "144",
    year = "1984",
pages = "37"
}

@article{Percacci_1991,  
	year = 1991,
	volume = {353},
	author = {R. Percacci},
	journal = {Nucl. Phys. B},
pages = "271"
}

@article{Nesti_2008,
	year = 2008,
	volume = {41},
	author = {F. Nesti and R. Percacci},
	journal = {J.Phys.A},
pages = "075405"
}

@article{Nesti_2010,
	year = 2010,
	volume = {81},
        author = {F. Nesti and R. Percacci},
	journal = {Phys. Rev. D},
   pages = "025010",
}

@article{Krasnov:2017epi,
    author = "Krasnov, K. and Percacci, R.",
    journal = "Class. Quant. Grav.",
    volume = "35",
    number = "14",
    year = "2018",
pages = "143001"
}

@article{Coleman1967,
  year = {1967},
  publisher = {American Physical Society ({APS})},
  volume = {159},
  author = {S. Coleman and J. Mandula},
  journal = {Phys. Rev.},
  pages = {1251},
}

@article{Alvarez_Gaume_2016,
	year = 2016,
	volume = {64},
	author = {L. Alvarez-Gaume and A. Kehagias and C. Kounnas and D. Lust and A. Riotto},
	journal = {Fortsch. Phys.},
pages = "176."
}

@article{D_Auria_2001,
		year = 2001,
	volume = {40},
	author = {R. D'Auria and S. Ferrara and M.A. Lled{\'{o}} and V.S. Varadarajan},
	journal = {J. Geom. Phys.},
        pages = {101},
}

@misc{majoranaspinors,
author = {J. Figueroa-O’Farrill},
url = {www.maths.ed.ac.uk/~jmf/Teaching/Lectures/Majorana.pdf},
}

@inproceedings{McKay1981TablesOD,
  booktitle={Tables of Dimensions, Indices, and Branching Rules for Representations of Simple Lie Algebras},
  author={W. McKay and J. Patera},
  year={1981}
}

@article{Polchinski:1984uw,
    author = "Polchinski, J.",
    journal = "Nucl. Phys. B",
    volume = "231",
    year = "1984",
    pages = {269},
}

@article{Manolakos:2023hif,
    author = "Manolakos, G. and Manousselis, P. and Roumelioti, D. and Stefas, S. and Zoupanos, G.",
journal = "Eur.Phys.J.ST",
doi = "https://doi.org/10.1140/epjs/s11734-023-00830-8",
year = "2023",
eprint = "2305.11785",
    archivePrefix = "arXiv",
}

@article{Feger_2020,
	year = 2020,
	volume = {257},
	author = {R. Feger. and T.W. Kephart and R.J. Saskowski},
	journal = {Comput. Phys. Commun.},
        pages = {107490},
}

@article{PhysRevD.31.1718,
  author = {Chang, D. and others},
  journal = {Phys. Rev. D},
  volume = {31},
  year = {1985},
  pages = {1718},
}

@article{PhysRevLett.52.1072,
  author = {Chang, D. and Mohapatra, R.N. and Parida, M.K.},
  journal = {Phys. Rev. Lett.},
  volume = {52},
  year = {1984},
  pages = {1072},
}

@article{Slansky:1981yr,
    author = "Slansky, R.",
    journal = "Phys. Rept.",
    volume = "79",

    year = "1981",
    pages = {1},
}

@article{Wilson:1973jj,
    author = "Wilson, K.G. and Kogut, John B.",
    journal = "Phys. Rept.",
    volume = "12",
    year = "1974",
    pages = {75},
}

@article{Wilson:1971bg,
    author = "Wilson, K.G.",
    journal = "Phys. Rev. B",
    volume = "4",
    year = "1971",
    pages = {3174}
}

@article{Kadanoff:1966wm,
    author = "Kadanoff, L.P.",
    journal = "Phys. Phys. Fiz.",
    volume = "2",
    year = "1966",
    pages = {263},
}

@book{Lust:1989tj,
    author = "Lust, D. and Theisen, S.",
    title = "{Lectures on String Theory}",
    volume = "346",
    year = "1989"
}
\end{document}